\documentclass[aps,prb,amsmath,amssymb,reprint,superscriptaddress,preprintnumbers,showpacs,intlimits]{revtex4-1}
\usepackage{bm,mathrsfs}
\usepackage[mathcal]{euscript}
\usepackage{hyperref}
\usepackage{graphicx}
\usepackage{subcaption}
\renewcommand{\vec}[1]{\bm{#1}}
\begin{document}

\title{Mechanism of Fast Axially--Symmetric Reversal of Magnetic Vortex Core}%
\author{Oleksandr V.~Pylypovskyi}
\affiliation{Taras Shevchenko National University of Kiev, 01601 Kiev, Ukraine}
\email{engraver@univ.net.ua}

\author{Denis D.~Sheka}
\affiliation{Taras Shevchenko National University of Kiev, 01601 Kiev, Ukraine}

\author{Volodymyr P.~Kravchuk}%
\affiliation{Institute for Theoretical Physics, 03680 Kiev, Ukraine}%

\author{Yuri Gaididei}%
\affiliation{Institute for Theoretical Physics, 03680 Kiev, Ukraine}%

\author{Franz G.~Mertens}%
\affiliation{Physics Institute, University of Bayreuth, 95447 Bayreuth, Germany}%

\date{December 10, 2012}




\begin{abstract}
The magnetic vortex core in a nanodot can be switched by an alternating transversal magnetic field. We propose a simple collective coordinate model which describes comprehensive vortex core dynamics, including resonant behavior, weakly nonlinear regimes, and reversal dynamics. A \emph{chaotic dynamics} of the vortex polarity is predicted. All analytical results were confirmed by micromagnetic simulations.
\end{abstract}


\maketitle

\section{Introduction}

Manipulation of complex magnetization configurations at the scales of nanometers and picoseconds is crucial for the physics of nanomagnetism\cite{Braun12}. Among the variety of different topologically nontrivial configurations special interest attracts the vortex configuration: it can form a ground state of the micro-- and nanosized disk--shaped particles (nanodisks). A magnetic vortex is characterized by an in--plane curling flux--closed structure, which minimizes the magnetostatic energy of the particle, and the out--of--plane region of the vortex core with about the size of the exchange length (typically about 10 nm for magnetically soft materials \cite{Wachowiak02}), which appears due to the dominant role of the exchange interaction inside the core \cite{Hubert98}. The direction of the vortex core magnetization, the so--called vortex polarity $p=\pm1$ (up or down), can be considered as a bit of information in the nonvolatile magnetic vortex random-access memories (VRAM)\cite{Kim08,Yu11a}. To realize the concept of VRAM one needs to control the vortex polarity switching process in a fast way.

There exist different ways to switch the vortex polarity. One can distinguish two basic scenarios of the switching: (i) Axially--asymmetric switching occurs, e.g. under the action of different in--plane AC magnetic fields or by a spin polarized current, see Ref.~\onlinecite{Gaididei08b} and references therein. Such a switching occurs due to the nonlinear resonance in the system of certain magnon modes with nonlinear coupling \cite{Kravchuk09,Gaididei10b}, which is accompanied by the temporary creation and annihilation of vortex--antivortex pairs. (ii) The axially--symmetric (or punch--through) switching occurs, e.g. under the influence of a DC transversal field  \cite{Okuno02,Thiaville03,Kravchuk07a,Vila09}. The mechanism of such a switching is the direct pumping of axially--symmetric magnon modes. Very recently the resonant pumping of such modes by an AC transversal field was proposed to switch the vortex in micromagnetic simulations \cite{Wang12,Yoo12}, which gives a possibility to achieve a switching at much lower field intensities.

The aim of the current study is to develop a theory for the axially--symmetric vortex polarity switching. We propose a simple analytical \emph{two--parameter cutoff model}, which allows to describe the main features of the complicated vortex dynamics under the action of AC pumping, including nonlinear resonance and magnetization reversal. Our model predicts a \emph{chaotic dynamics} of the vortex polarity, which is analyzed in terms of Poincar{\'e} maps. Our full--scale micromagnetic simulations confirmed all analytical predictions.

\section{Two--parameter cutoff model}
\label{sec:model}

We consider the model of a classical 2D Heisenberg ferromagnet with effective easy--plane anisotropy, caused by the dipolar interaction, under the action of a transversal AC field. The energy of such a magnet, normalized by the value $\pi A$ with $A$ being the exchange constant reads:
\begin{equation} \label{eq:energy}
\begin{split}
E &= \frac12 \int W \mathrm{d}^2x,\\
W &= \frac{(\vec\nabla m)^2}{1-m^2} + \left(1-m^2\right)(\vec\nabla \phi)^2 + \frac{m^2}{\ell^2}-\frac{2m h(\tau)}{\ell^2}.
\end{split}
\end{equation}
Here $m$ and $\phi$ are related to the components of the magnetization vector $\vec{M} = M_S\left( \sqrt{1-m^2}\cos\phi,  \sqrt{1-m^2}\sin\phi, m\right),$ the parameter $\ell=\sqrt{A/4\pi M_S^2}$ is the exchange length, $M_S$ is the saturation magnetization, and $h(\tau)= h\sin\omega \tau$ is the dimensionless external AC field. We use here the dimensionless time $\tau=\Omega_0 t$ with $\Omega_0 = 4\pi \gamma_0 M_S$ with $\gamma_0$ being the gyromagnetic ratio. The magnetization dynamics follows the Landau--Lifshitz equations, which can be derived from the following Lagrangian
\begin{equation} \label{eq:L}
L = G - E, \quad G = \frac{1}{2\pi\ell^2}\int \mathrm{d}^2x \left(1-m\right) \dot\phi
\end{equation}
and the dissipation function
\begin{equation} \label{eq:F}
F = \frac{\eta}{2\pi\ell^2}\int \mathrm{d}^2x \left[\frac{{\dot{m}}^2}{1-m^2} + (1-m^2){\dot{\phi}}^2\right].
\end{equation}
Here and below the overdot means the derivative with respect to $\tau$, the parameter $\eta$ is the Gilbert damping coefficient.

In order to describe the switching phenomena we propose a simple analytical picture using the following two--parameter Ansatz for the magnetization variables:
\begin{equation} \label{eq:ansatz}
m(r,\tau) = \mu(\tau) f\left(\frac{r}{\ell}\right), \; \phi(r,\tau) = \chi \pm\frac{\pi}{2} + \psi(\tau) g\left(\frac{r}{\ell}\right).
\end{equation}
We consider the vortex core amplitude $m(0,\tau)=\mu(\tau)$, which direction has the sense of the  \emph{dynamical vortex polarity} and is considered as a collective variable together with the in--plane \emph{turning phase} $\psi(\tau)$, the functions $f(x)$ and $g(x)$ describe the vortex structure. We use a Gaussian distribution for both functions, $f(x)=g(x)=\exp(-x^2/2)$, which is in a good agreement with simulation data. One has to note that such an Ansatz describes an axially--symmetric vortex solution together with the simplest axially--symmetric magnon mode: the real solution just slightly varies the profile of the functions $f(x)$ and $g(x)$. Also it is possible to take into account higher modes (with additional nodes on $r$), but we try to make the picture as simple as possible.

Using this Ansatz one can calculate the total energy of the vortex state disk as follows:
\begin{equation*}
E = \int_a^R W r\mathrm{d}r = \ln\frac{R}{a} + \mathscr{E}(\mu,\psi).
\end{equation*}
Here $R$ is the disk radius and $a$ is a cutoff parameter, which is of order of the magnetic lattice constant $a_0$. We introduce here the cutoff in order to take into account discreteness effects. It is worth to remind that in continuum theory two vortex states with different polarities are separated by an infinite barrier which prohibits the switching in a simply connected domain. The function $\mathscr{E}(\mu,\psi)$ is the effective energy of the system:

\begin{figure}
\begin{center}
\includegraphics[width=\columnwidth]{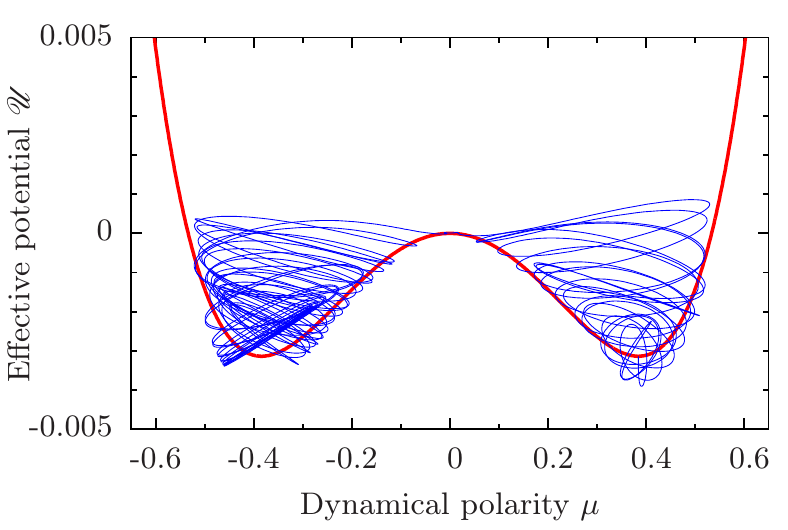}
\end{center}
\caption{Effective double--well potential $\mathscr{U}(\mu)$, see \eqref{eq:L(mu)} (solid line). The cutoff parameter $\lambda=0.07$. The typical evolution of the energy as function of the dynamical polarity $\mu(\tau)$ under the periodic pumping for unidirectional switching is shown by the thin line as a result of numerical integration of Eqs.\eqref{eq:EoM}. System parametersp: $h=0.001$ with frequency $\omega=0.7547$, damping $\eta=0.001$, initial conditions $\mu(\tau=0)=\mu_0 = 0.384$, $\psi(\tau=0)=0$, integration time $\tau_\text{all}=5000$, the trajectory is shown till $\tau_\text{sh}=500$.
}
\label{fig:en}
\end{figure}

\begin{equation} \label{eq:E(mu,psi)}
\begin{split}
\mathscr{E} &= \frac{\varkappa \mu^2}{2} + \frac12 \mathrm{Li}_2\left(\mu^2 \right) + \frac{\psi^2}{2}  - \frac{\mu^2 \psi^2}{8} - 2h \mu \sin\omega\tau.
\end{split}
\end{equation}
Here $\mathrm{Li}_2(x)$ is a dilogarithm function \cite{Abramowitz64}, $\varkappa = \ln\lambda+\gamma+1$ with $\gamma\approx0.577$ being Euler's constant, and we assume that $\lambda = a^2/\ell^2\ll 1$.

The effective Lagrangian and dissipation functions take the following forms:
\begin{equation} \label{eq:L-F(mu,psi)}
\begin{split}
\mathscr{L} &= -\frac12 \mu \dot{\psi} - \mathscr{E},\\
\mathscr{F} &= \frac{\eta}{2} \left[ -\frac{{\dot{\mu}}^2}{\mu^2} \ln\left(1-\mu^2\right) + {\dot{\psi}}^2 - \frac12 \mu^2{\dot{\psi}}^2 \right]
\end{split}
\end{equation}
The effective equations of motion are then obtained as Euler--Lagrange equations
\begin{equation} \label{eq:Euler-Lagrange}
\frac{\partial \mathscr{L}}{\partial X_i} -\frac{\mathrm{d}}{\mathrm{d} \tau} \frac{\partial \mathscr{L}}{\partial \dot{X}_i} =
\frac{\partial \mathscr{F}}{\partial \dot{X}_i}, \quad X_i = \left\{ \mu, \psi \right\},
\end{equation}
which finally read:
\begin{subequations} \label{eq:EoM}
\begin{align} \label{eq:EoMpsi}
\dot{\psi} &= - 2\varkappa \mu + \frac12\mu\psi^2 + \frac{2}{\mu} \ln(1-\mu^2) + 4h\sin\omega\tau\nonumber\\
& + 2\eta \frac{\dot{\mu}}{\mu^2}\ln(1-\mu^2),\\
\label{eq:EoMmu} %
\dot{\mu}  &= -\frac12\psi\left( \mu^2-4\right) - \eta (\mu^2-2)\dot{\psi}.
\end{align}
\end{subequations}

We start with the case without damping, $\eta=0$. In this case one can easily exclude the turning phase $\psi$ from the consideration, which results in the following effective Lagrangian for the dynamical polarity only:
\begin{equation} \label{eq:L(mu)}
\begin{split}
\mathscr{L}^{\text{ef}} &= \tfrac{1}{2}\mathscr{M}(\mu) \dot{\mu}^2 - \mathscr{U}(\mu) + 2\mu h\sin\omega\tau,\\
\mathscr{M}(\mu) &= \frac{1}{4-\mu^2}, \qquad \mathscr{U}(\mu) = \frac{\varkappa \mu^2}{2} + \frac12 \mathrm{Li}_2\left(\mu^2 \right).
\end{split}
\end{equation}
The Lagrangian \eqref{eq:L(mu)} describes the motion of a particle with the variable mass $\mathscr{M}(\mu)$ in the double--well potential $\mathscr{U}(\mu)$ under the action of periodical pumping. The typical shape of the potential $\mathscr{U}(\mu)$ is shown in Fig.~\ref{fig:en}: it has two energetically equivalent ground states with $\mu=\pm\mu_0$, which correspond to vortices with opposite polarities. In our cutoff model $\mu_0$ is a nonzero solution of the transcendent equation:
\begin{equation} \label{eq:mu0}
\varkappa \mu_0^2 = \ln\left(1-\mu_0^2\right).
\end{equation}
For $\lambda = 0.07$, the energy minimum corresponds to $\mu_0 = 0.384$, see Fig.~\ref{fig:en}. One has to note that our model works only for $|\mu|<1$. Another method is to work with a trigonometric variable, the vortex core angle $\vartheta$, instead of the dynamical polarity $\mu=\cos\vartheta$. We checked that the usage of $\vartheta$ provides the same physical picture, but the effective equations look awkward, so we keep to work with $\mu$.

The dynamical polarity $\mu(\tau)$ satisfies the following equation, see \eqref{eq:L(mu)}:
\begin{equation} \label{eq:ddotmuFull}
\ddot \mu + \mathscr{M} \mu\dot\mu^2 + \frac{\varkappa \mu}{\mathscr{M}} - \frac{\ln(1-\mu^2)}{\mathscr{M}\mu} = \frac{2 h}{\mathscr{M}} \sin\omega \tau.
\end{equation}
The linear oscillations near the potential well bottom have the usual harmonic shape:
\begin{equation} \label{eq:omega0}
\mu = \mu_0 + a e^{i\omega_0\tau}, \; \omega_0 = \sqrt{\frac{2\left(1+\varkappa -\varkappa \mu_0^2\right)}{\mathscr{M}_0(1-\mu_0^2)}}, \; |a|\ll1
\end{equation}
with $\mathscr{M}_0 = 1/(4-\mu_0^2)$ being the effective mass of a small oscillating particle near the well bottom and $\omega_0$ being the eigenfrequency.

\begin{figure}
\begin{center}
\includegraphics[width=\columnwidth]{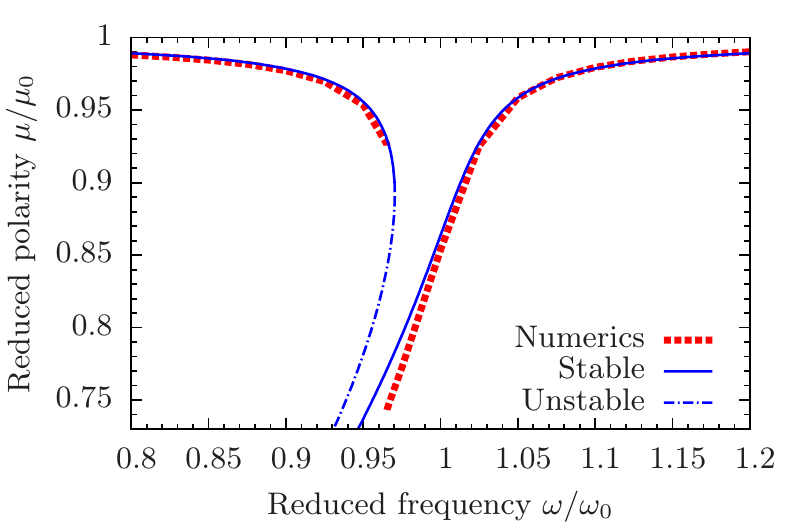}
\end{center}
\caption{The nonlinear resonance: amplitude-frequency characteristics from numerical solution of Eqs.~\eqref{eq:EoM}~(dashed curve) with initial conditions $\mu(\tau=0)=\mu_0$ and $\psi(\tau=0)=0$. The cutoff parameter $\lambda=0.07$, the damping $\eta=0.001$. The thin curves correspond to the analytical solution \eqref{eq:nonlinFreq} without damping: the solid lines correspond to the stable solution and the dash-and-dot line to the unstable region \eqref{eq:instability}. The field amplitude $h/\eta=0.15$. The total computation time for each point $\tau_\text{tot}= 3000$.
}
\label{fig:res}
\end{figure}

Let us study the weakly nonlinear dynamics using the method of multiple scales \cite{Kevorkian81,Nayfeh85,Nayfeh08}. We limit ourselves  by the three-scale expansion as follows:
\begin{equation} \label{eq:multscales}
\begin{split}
\mu    &= \mu_0 + \sum_{n=1}^3 \varepsilon^n \mu_n(T_0,T_1,T_2), \qquad T_n = \varepsilon^n t,\\
\omega &= \omega_0 + \varepsilon^2 \omega_2, \qquad h = \varepsilon^3 h_3\ll1,
\end{split}
\end{equation}
which provides a valid weakly nonlinear expansion under condition that the field amplitude is much less than the frequency detuning, $h/(\omega-\omega_0)\ll1$. Eq.~\eqref{eq:ddotmuFull} together with an expansion \eqref{eq:multscales} results in the set of equations for $\mu_n$, see Appendix~\ref{sec:multscale} for details. Following the Floquet theory \cite{Nayfeh85} one has to remove the mixed-secular terms in such equations, which finally provides the nonlinear resonant curve (see Appendix~\ref{sec:multscale} for details):
\begin{equation} \label{eq:nonlinFreq}
\omega_\pm (|a|)= \omega_0 - c_1 |a|^2 \pm c_2 \frac{h}{|a|},
\end{equation}
where $|a|$ is an amplitude of oscillations, see \eqref{eq:omega0}, the parameters $c_1$ and $c_2$ are calculated in \eqref{eq:eqforamp}. A typical nonlinear resonance curve is plotted in the Fig.~\ref{fig:res}. The low frequency branch $\omega_-$ contains a shock-stalling region. The upper limit of such an instability region can be found using the condition $\partial \omega_-/\partial a = 0$, which finally results in the limit frequency
\begin{equation} \label{eq:instability}
\omega_{\text{u}} = \omega_0 - 3\left(\frac{\sqrt{c_1}c_2h}{2}\right)^{2/3}.
\end{equation}

The unstable part of the resonance curve is plotted in Fig.~\ref{fig:res} by the dash-and-dot line. Further increase of the field amplitude leads to a broader instability domain. Moreover, we will see below that stronger pumping results in an essentially different kind of dynamics, leading to the switching of the vortex polarity and to chaotic behaviour.

\begin{figure}
\begin{center}
\includegraphics[width=\columnwidth]{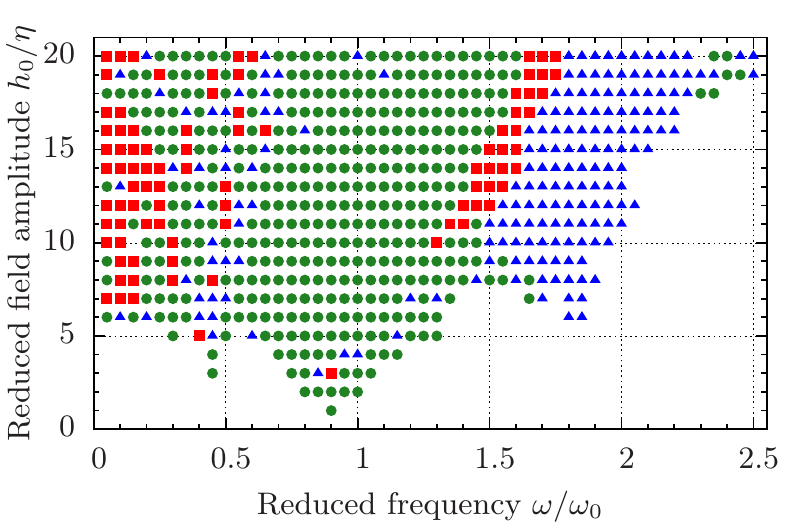}
\end{center}
\caption{Diagram of dynamical regimes for the field parameters (amplitude and frequency) as numerical solution of Eqs.~\eqref{eq:EoM}: circles correspond to one-period oscillations, triangles correspond to multiple-periods oscillations, and squares correspond to chaotic dynamics. White domains correspond to the absence of switching. Parameters are the same as in Fig.~\ref{fig:res} (the total computation time $\tau_\text{tot}$ depends on the regime, see the main text).
}
\label{fig:phd}
\end{figure}

If we increase the amplitude of the forcing, the system goes to the strongly nonlinear regime. We analyse such regimes using numerical solutions of Eqs.~\eqref{eq:EoM}. First of all, the regular oscillations of the dynamical polarity $\mu(\tau)$ between two potential wells occur in a wide range of parameters (the typical oscillations $\mathscr{U}(\mu)$ are plotted by the thin curve in Fig.~\ref{fig:en}). The diagram of dynamical regimes is shown in Fig.~\ref{fig:phd}. Different types of dynamical regimes are classified in accordance to the Poincar\'{e} maps. These maps are constructed for 20\,000 periods of the field oscillations for frequencies greater than $0.6\omega_0$ and for 15\,000 oscillation periods for other frequencies. The first 5000 points are dropped from consideration in order to exclude transient processes. The diagram of dynamical regimes has a resonant behaviour at the frequencies $\omega_0/3$, $\omega_0/2$, $\omega_0$ and $2\omega_0$, see Fig.~\ref{fig:phd}.

\begin{figure}

\begin{center}
\includegraphics[width=\columnwidth]{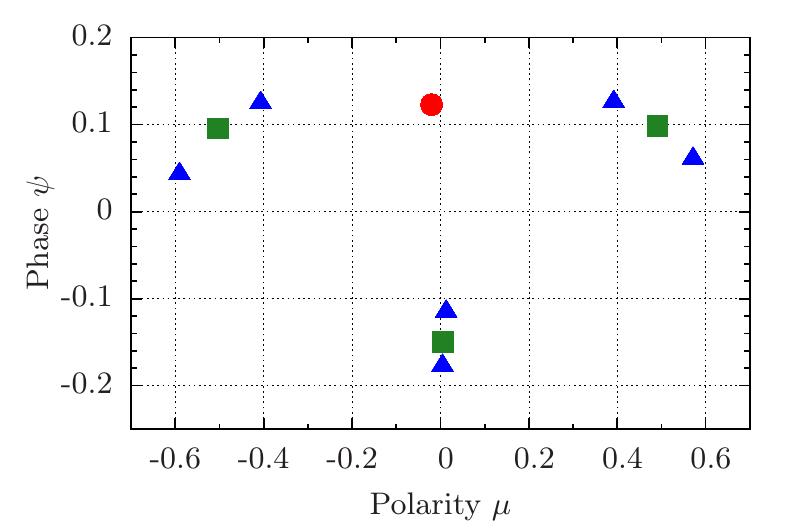}
\end{center}
\caption{Poincar\'{e} map for regular oscillations. The circle corresponds to $h/\eta =5$, $\omega/\omega_0=0.7$, triangles correspond to $h/\eta =11$, $\omega/\omega_0=2$, and squares correspond to $h/\eta = 13$, $\omega/\omega_0=1.95$.  Parameters are the same as in Fig.~\ref{fig:res}.
}
\label{fig:map3}
\end{figure}

There are three different dynamical regimes on the diagram: (i) The one-period oscillations (circles in  Fig.~\ref{fig:phd}) occur in a wide range of parameters, generally in the vicinity of the resonance frequency $\omega_0$, the Poincar\'{e} map for this regime has one stable focus (a circle in Fig.~\ref{fig:map3}). (ii) The multiple-period oscillations (triangles in Fig.~\ref{fig:phd}) occur typically near the doubled resonance frequency; the Poincar\'{e} map has a few points which are attended every pumping period and the trajectory in phase space $(\mu,\psi)$ makes a few windings before closing, see Fig.~\ref{fig:map3}.

\begin{figure}
\begin{subfigure}{\columnwidth}
\caption{$h/\eta=11$, $\omega/\omega_0 = 1.4$ , 15\,000 points}
\label{fig:Pmap1}%
\includegraphics[width=\columnwidth]{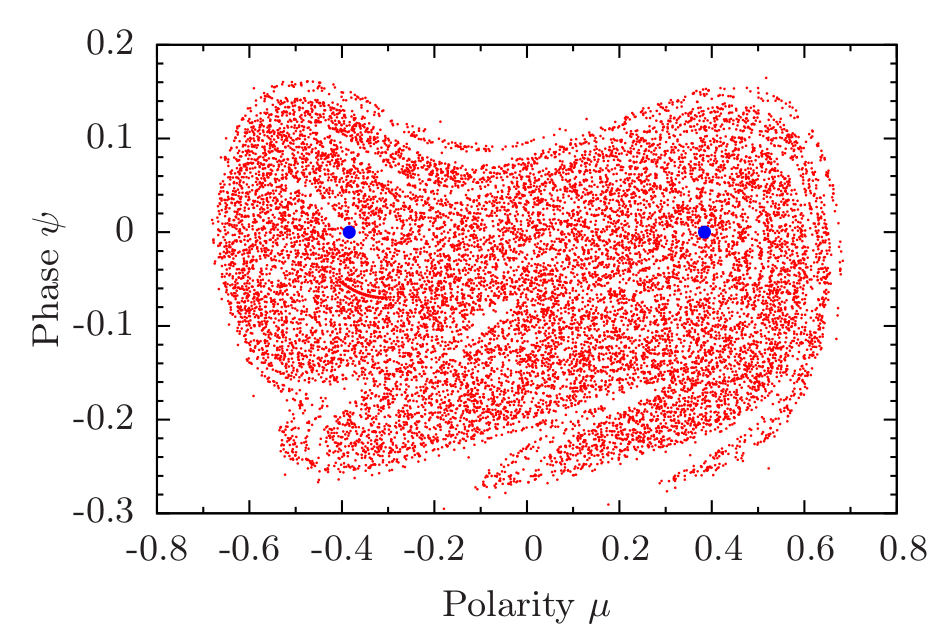}
\end{subfigure}%
\\ %
\begin{subfigure}{\columnwidth}
\caption{$h/\eta=15$, $\omega/\omega_0 = 0.1$, 10\,000 points}
\label{fig:Pmap2}%
\includegraphics[width=\columnwidth]{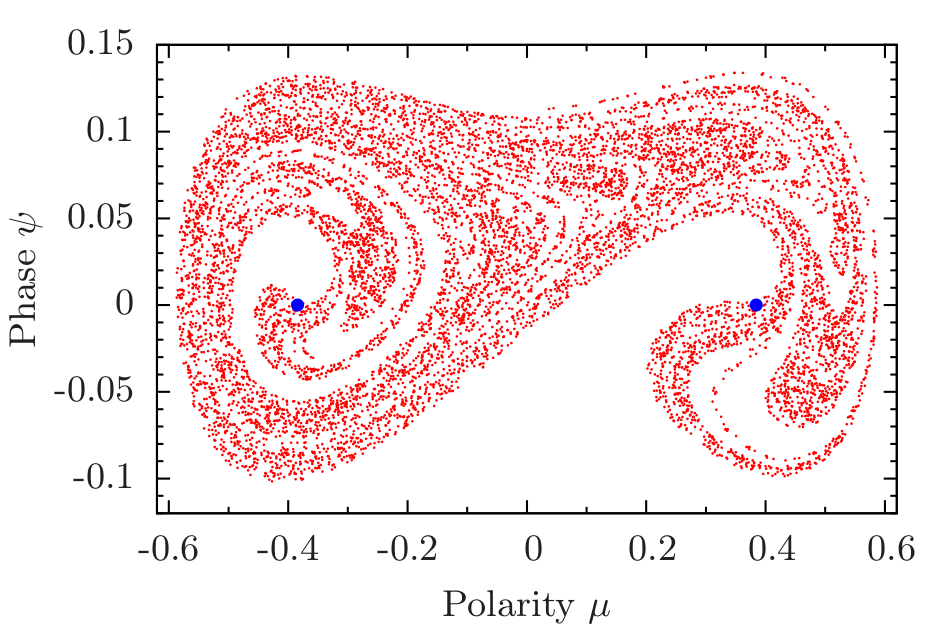}
\end{subfigure}%
\caption{Strange attractors on the Poincar{\'e} map. Equilibrium polarities are marked by the circles.
}
\label{fig:map}
\end{figure}

(iii) The chaotic oscillations of the dynamical polarity $\mu$ (squares in the Fig.~\ref{fig:phd}) take place in the transition region between the oscillations of the types (i) and (ii). The corresponding Poincar\'{e} map has the shape of a strange attractor, see Fig.~\ref{fig:Pmap1}. Apart them the chaotic dynamics occurs at the resonance frequency the weak enough field amplitude and in the wide range of the low frequency pumping. The low--frequency dynamics also corresponds to a strange attractor, a typical picture is presented in Fig.~\ref{fig:Pmap2}.

\section{Numerical study of the different dynamical regimes}
\label{sec:sim}

\begin{figure}
\begin{center}
	\includegraphics[width=\linewidth]{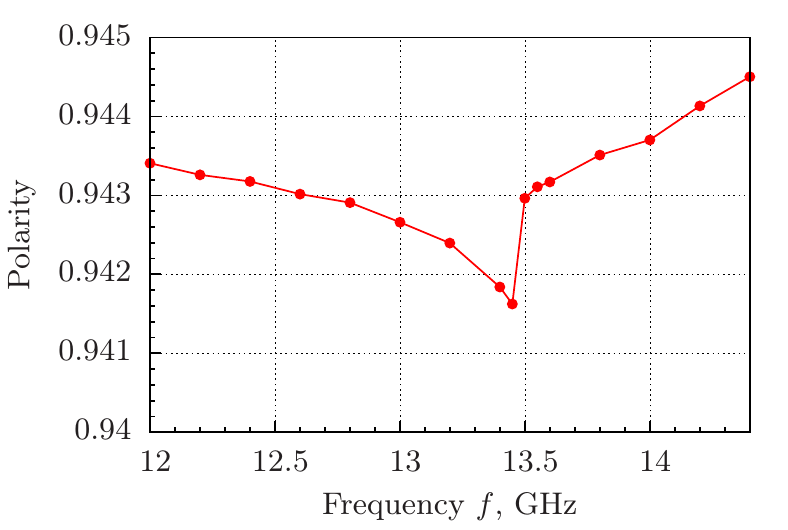}
\end{center}
\caption{The instability domain near the first resonance frequency for simulations with the field amplitude 5\,mT. The region near the first axially--symmetric harmonic is shown.}
\label{fig:instability}
\end{figure}

\begin{figure*}
\begin{center}
\includegraphics[width=\textwidth]{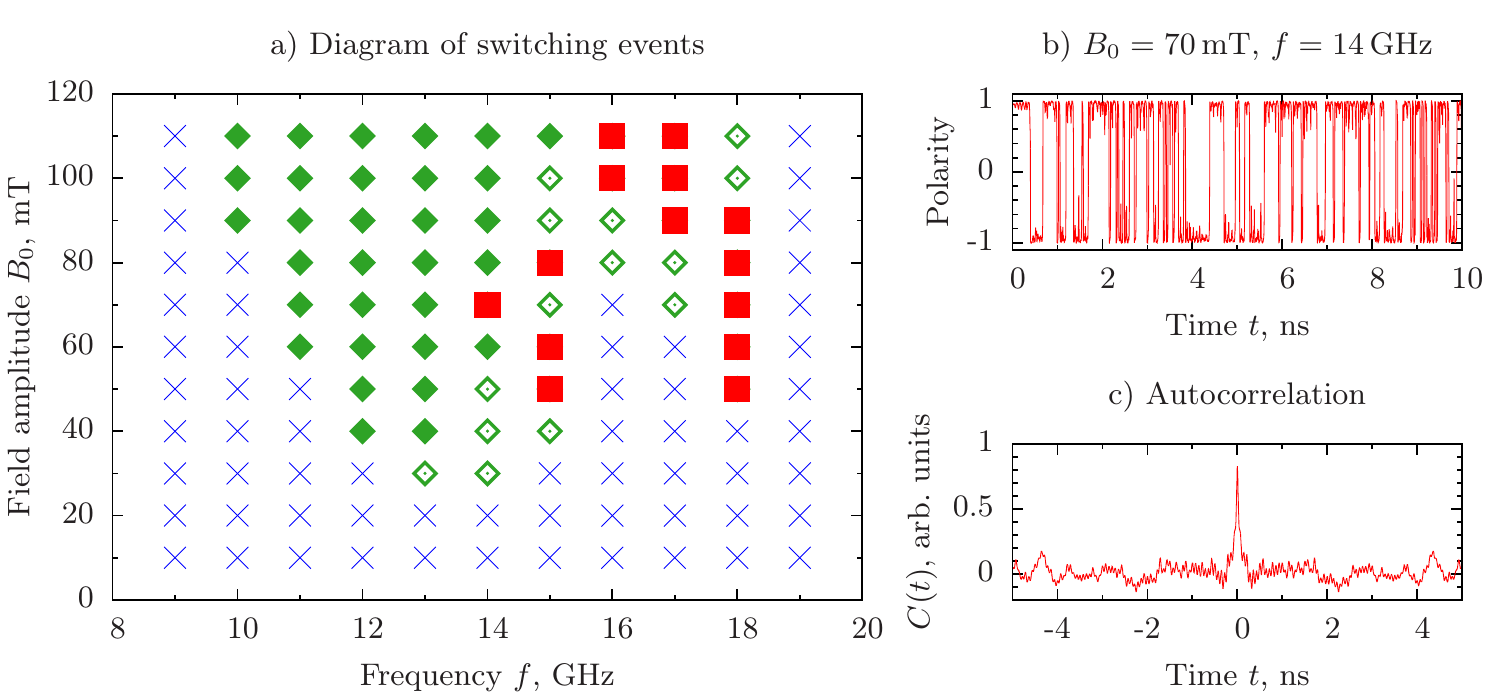}
\end{center}
\caption{Numerical simulations of the vortex dynamics: a) Diagram of dynamical regimes for the field parameters (amplitude and frequency). Diamonds and squares correspond to the switching events (the filled diamonds mark parameters, where the vortex rapidly moves away from the origin after a few switchings, the open diamonds mark the vortex which does not change its position); the filled squares show parameters for the chaotic polarity dynamics and the crosses correspond to the region without switching events. b) Example of the chaotic dynamics for the applied field amplitude $B=70$\,mT and the frequency $f=14$\,GHz. c) Autocorrelation function for the plot b).
}
\label{fig:swEvents}
\end{figure*}

In order to check all predictions of the two--parameter cutoff model, we performed a full--scale numerical modelling using the OOMMF framework~\cite{OOMMF}, which simulates the Landau--Lifshitz equations. Numerically we modelled a cylinder--shaped sample with radius $R=99$\,nm and height $L=21$\,nm using material parameters for Permalloy (Ni$_{80}$Fe$_{20}$): exchange constant $A=26$\,pJ/m, saturation magnetization $M_S=860$\,kA/m and Gilbert damping coefficient $\alpha=0.01$. The two--dimensional space mesh $3\times 3\times 21$\,nm is used. Initially, the vortex has an upward polarity and a counter--clockwise chirality.

The conditions of our numerical experiment were similar to simulations by \citet{Wang12} and \citet{Yoo12}. However, our task was to check the \emph{new} dynamical regime of \emph{chaotic dynamics}. That is why we need to study the long time dynamics.

First we examined the resonant frequency of radial spin waves. A 30\,mT constant pulse during 100\,ps was applied to the sample; it excited low--amplitude spin waves. By analysis of a Fourier spectrum for a 3.7\,ns long time dynamics of the total magnetization along the disk axis we calculated the eigenfrequency of the lowest spin wave mode equals $f_0=14$\,GHz.

We study the vortex dynamics under the action of the sinusoidal magnetic field
\begin{equation} \label{eq:fiels}
\vec B(t) = \vec{e_z} B \sin 2\pi ft,
\end{equation}
directed perpendicular to the face surfaces of the sample. The vortex dynamics under such a field has a resonant behaviour. The weak pumping causes the resonance on the frequency $f_0$. If we increase the field amplitude, the system goes to the nonlinear regime. The weakly nonlinear regime corresponds to the nonlinear resonance, see Fig.~\ref{fig:instability} (the resonance on the first axially--symmetric harmonic).

To systemize the complicated dynamics of the vortex polarity, we compute the phase diagram of the switching events by varying the field frequency in the vicinity of $f_0$ from 9 to 19\,GHz with steps of 1\,GHz and the field amplitude from 10 to 110\,mT, see Fig.~\ref{fig:swEvents}. There are two strong resonances in this range, which agree with previous results \cite{Wang12,Yoo12}. The lower resonance frequency is located between 12 and 14\,GHz; it corresponds to the axially symmetric mode without radial nodes ($m=0$, $n=0$). The second resonance is located near 18\,GHz; it corresponds to the axially symmetric mode with a single radial node ($m=0$, $n=1$). Since we expect a chaotic dynamics, it is necessary to analyse the long--time behaviour: numerically we checked the magnetization state every picosecond during 10\,ns interval. The chaotic vortex polarity dynamics during this time is observed in 14 simulations (see filled squares in the Fig.~\ref{fig:swEvents}a) where the vortex polarity switching mechanism corresponds to the axial--symmetric way. The typical shape of oscillations is presented in the Fig.~\ref{fig:swEvents}b) for $B=70$\,mT and $f=14$\,GHz. We examine the character of the polarity oscillations by an autocorrelation function
\begin{equation} \label{eq:autocorellation}
C(t_i) = \frac{1}{N}\sum_j \mu(t_{i+j}) \mu(t_j),\quad i,j=\overline{1,N},
\end{equation}
where the discretized time $t_n$ with steps $\Delta t=t_{n+1}-t_n =1$\,ps are used, and $N=10^4$ is a number of snapshots. The function $\mu(t_i)$ is the discrete dynamical polarity, which is defined as average magnetization of four cells in the center of the vortex core, normalized by the magnetization in the absence of the forcing. For the chaotic signal $C(t)$ rapidly decays, see Fig.~\ref{fig:swEvents}c). We marked points on the diagram of switching events by filled squares for simulations where the autocorrelation function rapidly decays and the distance between the maximum of the autocorrelation function and the first zero is smaller than 1\,ns. Plots of $C(t)$ for other simulations with  chaotic dynamics look similarly.

In all simulations the set of first switchings occurs during the first nanosecond and is accompanied by a high-amplitude axially--symmetric spin wave radiation. However, typically, the vortex position at the origin is unstable: during the field pumping the higher axially nonsymmetric modes ($m\neq0$) can be excited, which causes a vortex motion towards the disk edge surface. In such a case the switching occurs through the axially--\emph{asymmetric} mechanism, which is accompanied by the temporary creation and annihilation of a vortex-antivortex pair, see Ref.~\onlinecite{Gaididei08b} and references therein. Such  switching events are shown in the Fig.~\ref{fig:swEvents}a) by the filled diamonds. We do not analyse them due to an insufficiently short time interval, compared to the relaxation time, which corresponds to the axial--symmetric switching scenario, discussed in this work.

\section{Conclusions}

The axially--symmetric vortex polarity switching is an efficient way for the magnetization reversal on a subnanosecond time scale. Very recently such a scheme was realized by the micromagnetic simulations in Refs.~\onlinecite{Wang12,Yoo12}. To gain some insight to the resonant switching effect, \citet{Wang12} computed an exchange field inside the vortex core: it changes rapidly during the vortex reversal. \citet{Yoo12} noticed that the switching occurs only if the exchange energy exceeds a threshold value. The crucial role of the exchange interaction becomes clear in the analytical approach developed in the current study. Our two--parameter cutoff model explains the switching phenomenon in terms of the nonlinear resonance in a double--well potential. Such a potential arises mainly from the exchange interaction: the presence of two wells corresponds to the energy degeneracy with respect to the direction of the vortex polarity (up or down); the energy barrier between the wells becomes higher as the discreteness effects become less important.

In terms of our model the switching can be considered as the motion of an effective mechanical particle with a variable mass in the double-well potential. Under the action of periodical pumping the particle starts to oscillate near the bottom of one of the wells. When the pumping increases, there appear nonlinear oscillations of the particle; under a further forcing the particle overcomes the barrier, which corresponds to the magnetization reversal process. The chaotic dynamics of the magnetization is an analogue of the chaotic oscillations, e.g., in a Duffing oscillator\cite{Nayfeh08}.

In summary, we analyse analytically and numerically the axially--symmetric scenario of the vortex polarity switching, induced by an alternating magnetic field directed perpendicular to the nanodot surface. We propose a simple analytical two--parameter cutoff model, which describes the vortex polarity dynamics under such a resonance pumping and shows the possibility of both periodic and chaotic polarity oscillations by Poincar\'{e} maps. The micromagnetic simulations for Permalloy confirm a variety of the dynamical regimes and confirm our analytical predictions.

\begin{acknowledgments}
O.V.P. and D.D.S. thank the University of Bayreuth, where a part of this work was performed, for kind hospitality. O.V.P. acknowledges the support from the BAYHOST project. D.D.S. acknowledges the support from the Alexander von Humboldt Foundation.
\end{acknowledgments}

\appendix

\section{Analysis by the Method of Multiple Scales}
\label{sec:multscale}

We use the method of multiple scales\cite{Nayfeh85, Nayfeh08, Kevorkian81} to treat analytically Eq.~\eqref{eq:ddotmuFull}. We limit ourselves to the three--scale expansion \eqref{eq:multscales}. Since we have three different time scales $T_0$, $T_1$, and $T_2$, one has to modify the time derivatives as follows:
\begin{equation}
\frac{\mathrm{d}}{\mathrm{d}t} = \sum_{n=0}^2 \varepsilon^n D_n,\quad D_n = \frac{\mathrm{d}}{\mathrm{d}T_n}.
\end{equation}
The equations governing $\mu_1$, $\mu_2$, and $\mu_3$ are
\begin{subequations}
\begin{align} \label{eq:zeroorder}
&D_0^2 \mu_1 + \omega_0^2 \mu_1 = 0,\\
\label{eq:firstorder} %
&D_0^2 \mu_2 + \omega_0^2 \mu_2 = -[k_1 \mu_1^2 + k_{01}(D_0 \mu_1)^2+2D_0D_1\mu_1],\\
\label{eq:secorder} %
&D_0^2 \mu_3 + \omega_0^2 \mu_3 = -\Bigl\lbrace k_2 \mu_1^3 + D_1^2\mu_1 +\mu_1 (2 k_1 \mu_2 \nonumber\\
&+ k_{02} (D_0\mu_1)^2)+ 2 [k_{01} D_0\mu_1 (D_1\mu_1 + D_0\mu_2)\nonumber\\
&+ D_0D_2\mu_1+D_0D_1\mu_2]\Bigr\} + h_3 \sin(\omega_0T_0+\omega_2 T_2),
\end{align}
\end{subequations}
where we used the following notations:
\begin{align*}
k_{01}&= \mathscr{M}_0\mu_0,\qquad k_{02}= \mathscr{M}_0^2(4+\mu_0^2),\\
k_1&= -\frac{4-9 \mu_0^2-\mu_0^4+\varkappa  \left(1-\mu_0^2\right)^2 \left(4+3 \mu_0^2\right)}{\mu_0 \left(1-\mu_0^2\right)^2},\\
k_2&= -\frac{\mathscr{M}_0(12-27 \mu_0^2+38 \mu_0^4+\mu_0^6)+3 \varkappa  \left(1-\mu_0^2\right)^3}{3 \mu_0^2 \mathscr{M}_0 \left(1-\mu_0^2\right)^3}.
\end{align*}
The solution of the~Eq.~\eqref{eq:zeroorder} reads $\mu_1 = A(T_1,T_2)e^{i\omega_0T_0} + A^*(T_1,T_2)e^{-i\omega_0T_0}$.
To prevent the secular terms in the~Eq.~\eqref{eq:firstorder}, one has to put $A(T_1,T_2) \equiv A(T_2)$; the same condition for Eq.~\eqref{eq:secorder} gives an equation for the oscillation amplitude of $\mu_1$:
\begin{equation} \label{eq:eqforamp}
\begin{split}
&D_2A(T_2) + 4i c_1 A^2A^* = -h_3 \mathscr{M}_0 c_2 e^{i\omega_2 T_2},\\
&c_1 = \frac{5 k_{01} k_{1}}{12\omega_0} - \frac{3k_{2}}{8\omega_0} + \frac{5 k_{1}^2}{12\omega_0^3} + \frac{\omega_0}{24}\left(4 k_{01}^2-3 k_{02}\right),\\
&c_2 = \frac{1}{\mathscr{M}_0 \omega_0}.
\end{split}
\end{equation}
By solving the Eq.~\eqref{eq:eqforamp} one gets finally the~Eq.~\eqref{eq:nonlinFreq}.

%
%
%

\end{document}